\documentclass[twocolumn,11pt]{article}

\usepackage{epsfig}

\title {Experimental Implementation of the Quantum Baker's Map}
\author{Yaakov S. Weinstein$^\dag$, Seth Lloyd$^{*\sharp}$, Joseph V. Emerson$^\dag$, David G. Cory$^\dag$ \\
\\
$^\dag$\small{Department of Nuclear Engineering, M.I.T., Cambridge, MA 02139}
\\
$^*$ \small{d'Arbeloff Laboratory for Information Systems and Technology} \\
\small{Department of Mechanical Engineering, M.I.T., Cambridge, MA 02139}
\\
$^\sharp$\small{Author to whom correspondence should be addressed}
\\}

\begin{document}

\twocolumn[

\maketitle

\begin{abstract}
This paper reports on the experimental implementation of the quantum
baker's map via a three bit nuclear magnetic resonance (NMR) quantum 
information processor. 
The experiments tested the sensitivity of the quantum chaotic map to 
perturbations. In the first experiment, the map was iterated forward 
and then backwards to provide benchmarks for intrinsic errors and 
decoherence. In the second set of experiments, the least significant qubit 
was perturbed in between the iterations to test the sensitivity of
the quantum chaotic map to applied perturbations.  These experiments are 
used to investigate previous predicted properties of quantum 
chaotic dynamics.
\\
\\
PACS numbers 03.67.-a, 03.67.Lx, 05.45.-a
\\
\\
\end{abstract}]

Chaos is a phenomenon in which nonlinear dynamical systems exhibit
heightened sensitivity to small perturbations \cite{L+L} \cite{B1} \cite{B2} 
\cite{H1} \cite{H2} \cite{P8} \cite{P9} \cite{SC}.  
The study of chaos is computationally intensive.  When direct 
experiments are not available, 
computers can be used to simulate chaotic dynamics and to calculate the
effect of perturbations.   But when the chaotic system 
is quantum-mechanical, simulating its dynamics on a classical computer
is notoriously difficult: the computational complexity of the calculation
rises exponentially with the number of degrees of freedom of the simulated
quantum chaotic system and with the accuracy to which the simulation
is to take place \cite{H1} \cite{Feyn} \cite{Ll1}.  
If one can simulate quantum chaos on a quantum
computer, by contrast, the computational complexity rises only as
a small polynomial in the number of degrees of freedom and in the
accuracy \cite{Ll1} \cite{Brun1} \cite{Sch}.  
Consequently, quantum computation represents a potentially powerful 
technique for investigating quantum chaos.  This paper reports on
an experimental demonstration of a quantum chaotic map.

Small perturbations to the initial state of a classical chaotic system
typically lead to large changes in behavior.  Two states
that are initially close are driven apart at a rate governed by the 
positive Lyapunov exponents of the chaotic dynamics.  By contrast,  
quantum dynamics, whether regular or chaotic, preserves the overlap between
quantum states and does not drive them apart. Nonetheless, quantum chaos
can be characterized by the sensitivity of the time evolution of states
to small changes in the Hamiltonian that governs the chaotic dynamics
\cite{P8} \cite{P9} \cite{SC} \cite{ZP1} \cite{B} \cite{Baranger}.

Peres \cite{P8} \cite{P9} noted that under quantum chaotic dynamics with a
slight perturbation, a state moves apart at an exponential rate from
the same state evolving under the unperturbed dynamics. Schack and Caves 
\cite{SC} characterized chaotic dynamics (both classical and quantum) in 
terms of the exponential growth of the information required to specify 
a state that evolves according to a perturbed version of the dynamics, a
phenomenon they termed `hypersensitivity to perturbation.'
Zurek and Paz \cite{ZP1} conjectured that quantum chaotic systems 
are sensitive to
weak interactions with the environment, which cause such systems to
produce information at a rate equal to their Kolmogorov-Sinai entropy. 
Ballentine and Zibin \cite{B} compare quantum and classical
chaotic systems by measuring the accuracy with which the system returns
to its initial state under time reversal after a perturbation of varying 
strength. The implementation of the baker's map reported here allows for the 
experimental investigation of quantum chaotic dynamics, specifically its 
sensitivity to perturbations as described by the above models.

The quantum information processor used to simulate perturbed quantum 
chaotic maps is an NMR quantum information processor \cite{Cory1} \cite{Gersh}.
The number of quantum bits used (three) is sufficiently small that
the precision of the quantum computation can be checked on a classical
computer.  Of course, the small number of qubits means that the simulation
could have been performed on a classical computer.  The goal of the
research reported here was to actually simulate quantum chaos on a quantum
information processor  
(as opposed to using a classical computer to simulate a quantum 
information processor simulating quantum chaos \cite{Brun1}).
We also explore techniques and problems for scaling up such
simulations to allow for a thorough investigation
of quantum chaotic dynamics of high-dimensional quantum systems.
A further goal of experimental investigations is to identify practical
experimental signatures of quantum chaos.

The quantum chaotic map investigated is the quantum baker's map 
\cite{B+V} \cite{sar}.
The classical baker's map acts on the unit square in phase 
space as follows:  The baker's map first stretches phase space to 
twice its length, 
while squeezing it to half its height. Then, the map cuts phase space 
in half vertically and stacks the right portion on top of the left portion, 
similar to the way a baker kneads dough. Because of the stretching and
the cut, the baker's map is fully chaotic and has two Lyapunov exponents, 
$\pm \ln 2$. Balazs and Voros \cite{B+V} presented a quantized version of
the baker's map that reproduces the behavior of the classical map in
the limit $\hbar \rightarrow 0$. The quantum baker's map (QB) is a simple 
unitary operator which consists of a quantum Fourier transform (QFT) on half 
of the Hilbert space followed by an inverse QFT on the whole Hilbert space. 

The quantum baker's map can be expressed as a sequence of two basic unitary
operations, the Hadamard gate, $H_j$, operating on spin $j$ and the 
conditional phase gate, $B_{jk}$, operating on spins $j$ and $k$. 
Schack \cite{Sch} utilized this decomposition to develop an algorithm for 
simulating the quantum baker's map on a quantum computer. The three qubit 
version of the quantum baker's map is
\begin{equation} 
Swap_{13}H_3B^\dag_{23}B^\dag_{13}H_2B^\dag_{12}H_1Swap_{23}H_3B_{23}H_2
\end{equation}
where $Swap_{jk}$ is a swap gate between bits $j$ and $k$. This gate sequence 
is a QFT on qubits 2 and 3 \cite{Cop}, followed by an inverse QFT on all 
three qubits.

We note that according to a well known conjecture 
\cite{B1}\cite{BGS}\cite{H1}, 
the energy eigenvalues of a quantum system  
with classically chaotic dynamics are expected to have the statistical 
properties of a Gaussian random matrix. A quantum baker's 
map whose Hilbert space dimension is a power of two (which is the case for our 
implementation) does not exhibit this spectral distribution \cite{B+V}. 
Nevertheless, it does show interesting behavior due to perturbations.

The approach taken in this implementation of the quantum baker's map is to
apply a perturbation between forward and backward map iterations.
In general, the final state with perturbation can be expressed as  
$\rho_f' = \sum_k(QB)^{\dag}A_k(QB)\rho_{initial}(QB)^{\dag}A_k^\dag(QB)$,
whereas the final state without perturbation is 
$\rho_f = (QB)^{\dag}(QB)\rho_{initial}(QB)^{\dag}(QB)$ In the absence of 
experimental errors $\rho_f = \rho_initial$.

An interesting non-unitary
perturbation is the application of a magnetic field gradient dephasing the 
least significant (third) bit. The Hamiltonian of a gradient on spin $j$ is 
$H_{g}^{j}(z) = e^{i\gamma\frac{dB_z}{dz}z\frac{\sigma_{z}^{j}}{2}}$, 
and it acts as a rotation of varying magnitude  across the sample. 
This perturbation looks like decoherence when measuring the signal from the
entire sample, effectively tracing over position, $z$.

An interesting type of unitary perturbation is one with a simple definition
in phase space, such as a displacement in position. The quantum
baker's map, as defined by equation (2), associates the discretized position 
basis on the unit square \cite{B+V} with the computational basis.
The smallest possible position perturbation in a Hilbert
space of dimension $N$ corresponds to the addition or subtraction of 
$1 (mod N)$  which, for 
example, would take the state $|000\rangle$ to $|001\rangle$. This perturbation
can be implemented on $n$ qubits by applying a $(j-1) \times$ controlled-NOT 
to 
bit $j$, where the controls are all bits less significant then $j$, as $j$ is 
indexed from $n$ to 1. Using geometric algebra \cite{Shy} this sequence for 
three bits can be written as follows:
\begin{eqnarray}
\begin{array}{c}
e^{i\frac{\pi}{8}(1-\sigma^1_x-\sigma^2_z-\sigma^3_z+\sigma^1_x\sigma^2_z+
\sigma^1_x\sigma^3_z+\sigma^3_z\sigma^3_z-\sigma^1_x\sigma^2_z\sigma^3_z)}
\times \\
e^{i\frac{\pi}{4}(1-\sigma^2_x-\sigma^3_z+\sigma^2_x\sigma^3_z)}
e^{i\frac{\pi}{2}(1-\sigma^3_x)}
\end{array}
\end{eqnarray}
Simulated data in the inset of figure (1) shows that, as expected, 
the baker's map is very sensitive to such a perturbation.

A variation of this perturbation with a simple interpretation in Hilbert 
space is an $x$ rotation on one bit. The sensitivity of the baker's map to 
the $\sigma_x$ perturbation may be carefully examined 
experimentally by controlling the angle of the rotation.
In our experiments we study only one initial state, the $|000\rangle$ 
psuedo-pure state. In general, the perturbed dynamics of a single state may 
not be sufficient to completely characterize the behavior of a map. 
Here, we take advantage of the fact that the overlap $Trace(\rho_f\rho_f')$ 
for the $|000\rangle$ initial state approaches the 
average of overlaps for a complete set of orthogonal initial states 
in the limit of small perturbation, as shown in figure (1).
Thus the overlap, $Trace(\rho_f\rho_f')$ for the initial state $|000\rangle$
approaches the fidelity \cite{Evan} \cite{C} of the operation: baker's map, 
rotation perturbation, inverse baker's map.

\begin{figure}
\epsfig{file=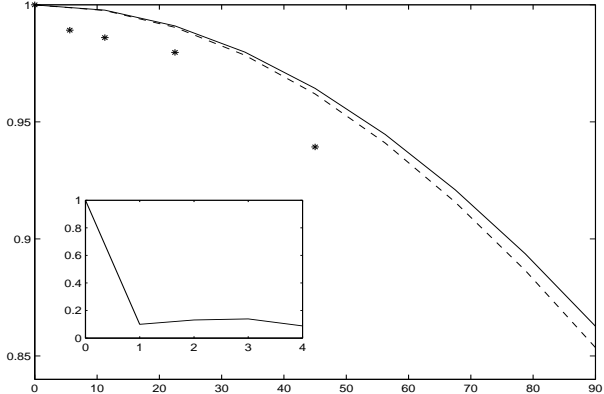, height=5.8cm, width=8cm}
\label{plot}
\caption {Sensitivity of baker's map to unitary perturbations. Solid line 
shows the theoretical overlap $Trace(\rho_f\rho_f')$ 
versus the angle of the perturbation
rotation on the third bit for the initial state $|000\rangle$. 
The dashed curve shows the final state overlap for the same 
perturbation averaged over a complete set 
of orthonormal initial states. For small perturbations the results for 
the initial state $|000\rangle$ are very similar to that of a complete 
set of initial states. 
Stars represent the measured overlap $Trace(\varrho_f\varrho_f')$
for the initial state $|000\rangle$. 
The inset plot is the overlap  
versus the size of the position perturbation (add 1 through  
add 4). The baker's map is much more sensitive to 
a position space perturbation then to a 
rotation perturbation on the least significant bit. }
\end{figure}

The three bit quantum baker's map was implemented via NMR using the 
three carbon-13 spins of an 
alanine sample. The resonant frequency of carbon-13 on a 300MHZ spectrometer 
is approximately 75.468MHz. Frequency differences between the spins are 
9456.5Hz between spins 1 and 2, 2594.3Hz between spins 2 and 3, and 12050.8Hz 
between spins 1 and 3. Coupling constants between the three spins are $J_{12}$
= 54.2Hz, $J_{23}$ = 35.1Hz, and $J_{13}$ = -1.2Hz. Relaxation time $T_1$ for 
the three carbon spins in alanine are all longer than 1.5s while the $T_2$ 
relaxation times are longer than 400ms.  The pulse sequences for realizing the
$H_j$ and $B_{jk}$ gates and the implementation of the QFT as a sequence of 
these gates, can be found in \cite{YSW}. The pulse sequence for the complete 
quantum baker's map was
compressed by relabeling bits instead of performing the swaps explicitly.
Readout was done using quantum state tomography as described in \cite{Evan}. 

To measure the accuracy with which the transformations were performed
we used the correlation measure introduced in \cite{C} which is appropriate 
for almost fully mixed density matrices: 
\begin{equation}
C= \frac{Tr(\varrho_{theory}\varrho_{exp})}{\sqrt{Tr(\varrho_{theory}^{2})}\sqrt
{Tr(\varrho_{exp}^{2})}}\sqrt{\frac{Tr(\varrho_{exp}^2)}{Tr(\varrho_{initial}^2)
}}.
\end{equation}
Here, $\varrho$ is the measured deviation density matrix. This does not 
include the large identity term in the actual high temperature liquid state,
but does include a small amount of the identity operator necessary for 
reconstructing a positive state operator. The amount of identity included is 
fixed for each set of experiments. If the theoretical and 
experimental deviation density matrices were correlated $C=1$, if they were 
uncorrelated $C=0$ and if they were anti-correlated $C=-1$. 

The first experiment consisted of two iterations of the quantum baker's map, a 
forward iteration followed by the inverse of the map, starting from the 
$|000\rangle$ pseudo-pure state. The attenuated correlation, $C$, of the 
implementation of the forward map is .76, and for the forward followed by the 
inverse, .56. The first term in the attenuated correlation 
measures the correlation between theoretical and experimental density 
matrices without accounting for reduction in signal over the course of the 
experiment. This gives a crude measure the maps' accuracy in the absence of 
decoherence. For the forward map this unattenuated correlation is .93 and for 
the forward followed by the inverse, .90. Since the experiment was done on the 
$|000\rangle$ pseudo-pure state, we expect the final state of the system to 
be that same state. The density matrix of the spin system after the forward 
and inverse iteration is shown in figure (2). 

\begin{figure}
\epsfig{file=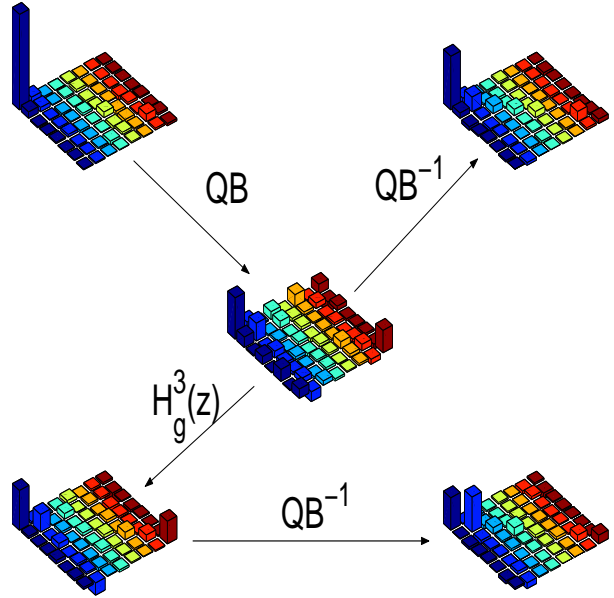, height=8cm, width=8cm}
\label{d}
\caption {Real part of experimental density matrices. The top left hand 
corner shows the
input $|000\rangle$ psuedo-pure state. One iteration of the baker's map
on this state leads to the center density matrix. Then the inverse map is 
applied (top right figure) bringing the bits  back to the input $|000\rangle$ 
state. In the second experiment, a  gradient is applied to the third, 
least significant, bit after the forward iteration of the map (bottom left 
figure) followed by the application of the inverse map (bottom right figure).}
\end{figure}

Another set of experiments were done to explore the dynamics of the baker's 
map under perturbations as described above. The first perturbation 
experimentally tested consisted of 
a dephasing gradient on the third, least significant, bit. We measured 
$C = .78$ for the state after the gradient and 
$C = .65$ for the state after the inverse map.
For these correlation values the loss of magnetization due to the 
gradient is taken into account in the normalization of $\varrho_{initial}$. 
The final state (after the gradient and inverse map) shows the 
generation of one bit of entropy as seen by the equilibration of the 
$|000\rangle$ and $|001\rangle$ populations, as displayed in figure 
(2). This is consistent 
with the Paz-Zurek model for the effect of decoherence on a quantum chaotic 
map: when decohered, the map produces one bit of entropy per iteration, an 
amount equal to the Kolmogorov-Sinai entropy of the map. 

In a final set of experiments rotational perturbations of $\pi/32$, $\pi/16$, 
$\pi/8$, and $\pi/4$ on the third (least significant) bit were applied 
between the forward and inverse iterations of the baker's map. 
For these experiments $C$ ranged between .52 and .53. In figure (1) 
we plot the overlap, $Trace(\varrho_{f}\varrho_{f}')$, of the experimental 
perturbed and unperturbed density matrices and compare it to 
theoretical predictions.   

In conclusion, we describe the implementation of a chaotic map 
on a quantum system, a three qubit quantum information processor. In addition,
we have explored two perturbations and examined their effects on the 
dynamics of the map. Experiments such as these establish a foundation for 
further experimental investigations of quantum chaotic dynamics 
and the exploration of suggested theoretical approaches. For example, 
the hypersensitivity to perturbation suggested by Schack
and Caves, should be evident even on a small Hilbert space, with 
only a few iterations of a chaotic map \cite{Brun1}. To eventually 
observe a characteristic such as the Peres criterion \cite{P8}
will require a much larger Hilbert space and many more iterations of the map. 
We believe the experiments performed are a first step towards a more thorough
experimental investigation of these questions. 

The authors thank M.A. Pravia and E.M. Fortunato for help with experimental
difficulties and J.P. Paz for helpful discussions. This work was 
supported by DARPA/MTO through ARO grant DAAG55-97-1-0342.

\end{document}